\documentclass[11pt]{article}
\usepackage{graphicx}

\newsavebox{\hflrar}
\sbox{\hflrar}{\makebox[0pt][l]
{${\scriptstyle \leftharpoonup}$}{${\scriptstyle \rightharpoonup}$}}

\def \to {\rightarrow}

\begin{document}
\begin{center}
{\Large\bf Transverse Momentum Dependent Factorization for Radiative Leptonic Decay  of $B$-Meson} \vskip 10mm
J.P. Ma and Q. Wang    \\
{\small {\it Institute of Theoretical Physics, Academia Sinica,
Beijing 100080, China }} \\
\end{center}

\vskip 1cm
\begin{abstract}
With a consistent definition of transverse-momentum-dependent(TMD) light-cone wave function of $B$-meson,
we show that the amplitude of the radiative leptonic $B$-decay  can be factorized at one-loop level
as a convolution with the wave function and a perturbative coefficient function, combined
with a soft factor. In this TMD factorization, the transverse momenta of partons in the $B$-meson
are taken into account and all soft divergences are contained in the wave function and the soft factor.
The coefficient function is infrared-safe. The factorization works on a diagram-by-diagram
basis and is possible to extend beyond one loop. With the factorization the
large logarithms in the perturbative function can be simply resummed.
Our work shows that the result of collinear factorization for the decay
can be derived from that of TMD factorization. Therefore, the two factorizations for the case here
are simply related to each other.
\vskip 5mm
\noindent
\end{abstract}
\vskip 1cm
\par
\eject
\noindent
{\bf\large 1. Introduction}
\par\vskip15pt
Exclusive B-decays play an important role for testing the standard
model and seeking for new physics. Experimentally they are studied
intensively. Theoretically, there are two approaches of QCD
factorization for studying these decays. One is based on the
collinear factorization\cite{BBNS}, in which the transverse momenta
of partons in a B-meson are integrated out and their effect at
leading twist is neglected. The collinear factorization has been
proposed for other exclusive processes for long time\cite{BL}.
Another one is based on $k_T$-factorization\cite{KT1} or pQCD approach,
where one takes the transverse momenta $k_T$ of
partons into account at leading twist by means of $k_T$-dependent  light-cone
wave function. We will call such a factorization as transverse momentum dependent(TMD) factorization.
The advantage of the TMD factorization is that it may
eliminate end-point singularities in collinear
factorization\cite{EPS} and some higher-twist effects are included.
The knowledge of the transverse momentum dependent(TMD) light-cone wave function
will provide a
3-dimensional picture of a $B$-meson bound state. However, it was
not clear how to define the TMD light-cone wave function in a
consistent way to perform a TMD factorization because of light-cone
singularities\cite{Col1}.
\par
Similar problems also appear
in defining TMD parton distributions and fragmentation functions
if one tries to do TMD factorization for inclusive processes.
In general the light-cone singularities appear if a parton
emits gluons carrying momenta which are vanishingly small in the $+$-direction
but large in other directions in a light-cone coordinate system. In a collinear factorization
for an exclusive or inclusive process, these singularities are cancelled
between different contributions if the transverse momentum of the parton
is integrated out. If the transverse momentum is not integrated, the singularities
are not cancelled.
\par
For inclusive processes like Drell-Yan, semi-inclusive DIS etc.,
it has been shown that one can consistently define TMD parton distributions
by using gauge links in the direction off the
light-cone direction and the TMD factorization of inclusive processes
can be done without light-cone singularities\cite{CS,CSS,JMY,JMYG}.
The TMD parton distributions defined with these gauge links
will depend on the deviation of the direction from the light-cone direction.
The evolution
with this dependence is controlled by the Collins-Soper equation\cite{CS} which
leads to the so-called CSS resummation formalism\cite{CSS,JMY,JMYG}. This formalism
is for resummation of large logarithms appearing in the collinear factorization.
In that sense the TMD- and collinear factorization are related to each other.
But the similar relation in exclusive B-decays has not been studied.
\par
We have proposed in \cite{MW} to consistently define the TMD light-cone wave function
of B-meson by using gauge links off the light-cone direction and studied its relation
to the usual light-cone wave function in the collinear factorization.
With the consistent definition it is important to show that the TMD factorization
can be consistently performed. The relation between two factorization approaches
may be then established. As a first step towards these goals
we study in this paper TMD factorization for the radiative leptonic decay of B-meson.
We also plan to study  TMD factorization for $B\to\pi$ form factor and other decay processes.
It should be noted that the definition of the TMD light-cone wave function
is not unique, different definitions are possible. A different definition
can be found in \cite{LiLiao}. With different definitions
the most important thing is to show that one can perform TMD
factorization consistently with one of these definitions, at least at one-loop level.
To our knowledge, there is so far no such a study beyond tree-level for exclusive B-decays.
We will show
that with the definition given in \cite{MW} the factorization
can be done at one-loop level for the process studied in this paper.
\par
The radiative leptonic $B$-decay has been studied extensively\cite{RLB1,KPY,DS,LPW,BHLN, Li3}.
The effect of strong interaction in the decay is parameterized with form factors.
These form factors have been studied in \cite{KPY,DS,LPW,BHLN, NaLi,Li3} with QCD
factorization.
It has been
shown that the form factors can be factorized as a convolution with a perturbative
coefficient function and the light-cone wave function of the $B$-meson
in the collinear factorization\cite{DS,LPW,BHLN,Li3}. In these works the transverse momentum
of partons is integrated out. It results in the convolution only with the $+$-component
of the parton momentum. In \cite{KPY} the transverse momentum of the parton
is not integrated and is explicitly taken into account, but the consistency
of the definition of the TMD light-cone wave function is not addressed and
the problem of the gauge invariance of the definition is ignored.
In \cite{NaLi} the decay is studied with $k_T$-factorization or TMD factorization,
but the TMD light-cone wave function employed there has the light-cone singularity.
\par
With our gauge-invariant
definition we can show with TMD factorization that the form factors take the factorized
form:
\begin{equation}
     \phi_+ \otimes \tilde S \otimes H.
\end{equation}
In the above $\phi_+$ is the TMD light-cone wave function, $\tilde S$ is a soft factor, $H$
is a coefficient function which can be calculated with perturbative QCD and is free
from soft divergences. $\phi_+$ and $\tilde S$ are well-defined matrix elements
of QCD operators. The convolution here is not only with  $+$-components  but also
transverse components of parton momenta. In this paper we prove the factorization at one-loop level.
We show that the cancellation of all soft divergences is on a diagram-by-diagram basis.
This is important for extending our factorization beyond one-loop level.
In the case studied in the paper,
the TMD factorization is similar to the collinear factorization because
there is no parton or hadron in the final state. But it is important to show
first that the TMD factorization works for this simple case and then extend
the TMD factorization to other cases. An interesting fact with the TMD factorization
is that it provides a simple way to resum large logarithms in the perturbative function
$H$, as we will show in the paper.
\par
As mentioned before, TMD factorization for an inclusive process can be related to
the corresponding collinear factorization. One can expect that such a relation also
exists for exclusive $B$-decays. Indeed, in the case studied here, such
a relation exists and it is simple: Both
factorizations are equivalent, i.e., one can derive the result of the collinear
factorization from our TMD factorization. We will show this in this work.
One reason for this simple relation is that there is no hadron, hence any parton
in the final state.
\par
Our paper is organized as the following: In the next section we define
the TMD light-cone wave function and give its one-loop result in detail
with a general partonic state, which will be used to perform
TMD factorization. A detailed discussion about the TMD light-cone wave function and
its relation to the usual light-cone wave function in collinear factorization
can be found in \cite{MW}.  In Sect. 3 we introduce our notation for the decay
and the result of the factorization at tree-level. In Sect. 4
we will complete the factorization at one-loop level and determine the soft factor.
In Sect.5 we show that the result of the collinear
factorization can be derived from that of the TMD factorization and establish
the relation between the two factorizations for the decay.
In Sect.6 we will make an attempt to re-sum large logarithms in TMD factorization.
Sect.7 is our conclusion and outlook.
\par
\vskip20pt
\noindent
{\bf\large 2. A Consistent Definition of the TMD Light-Cone Wave-Function}
\par\vskip15pt
In this section we give our definition of the TMD light-cone wave function
and its one-loop result in detail with a general partonic state.
A brief report of the result and the study of the relation
to the light-cone wave function in the collinear factorization can be found
in \cite{MW}.
\par
We will use the  light-cone coordinate system, in which a vector
$a^\mu$
is expressed as $a^\mu = (a^+, a^-, \vec a_\perp) = ((a^0+a^3)/\sqrt{2},
(a^0-a^3)/\sqrt{2}, a^1, a^2)$ and $a_\perp^2 =(a^1)^2+(a^2)^2$.
For b-quark we will use the heavy quark effective theory(HQET).
To define the TMD light-cone wave function
we introduce a vector $u^\mu=(u^+,u^-,0,0)$ and the definition is given
in the limit $u^+ << u^-$\cite{MW}:
\begin{eqnarray}
 \phi_+(k^+, k_\perp,\zeta, \mu) =\ \int \frac{ d z^- }{2\pi}
  \frac {d^2 z_\perp}{(2\pi )^2}  e^{ik^+z^- - i \vec z_\perp\cdot \vec k_\perp}
 \langle 0 \vert \bar q(z) L_u^\dagger (\infty, z)
  \gamma^+ \gamma_5 L_u (\infty,0) h(0) \vert \bar B(v) \rangle\vert_{z^+=0},
\end{eqnarray}
where $h(x)$ is the $b$-quark field in HQET.
and $L_u$ is the gauge link in the direction $u$:
\begin{equation}
L_u (\infty, z) = P \exp \left ( -i g_s \int_{0} ^{\infty} d\lambda
     u\cdot G (\lambda u+ z ) \right ) .
\end{equation}
In the above, the $B$-meson moves with the velocity $v^\mu =(v^+, v^-, 0,0)$,
i.e., in the $z$-direction.
The limit should be understood that we do not take the contributions
proportional to any positive power of $u^+/u^-$ into account.
This definition is gauge invariant in any non-singular gauge
in which the gauge field is zero at infinite space-time.
It has not the mentioned light-cone singularity as we will show through
our one-loop result, but it has an extra dependence
on the momentum $k^+$ through the variable
$\zeta^2 = 4(u\cdot k)^2/u^2=\zeta^2_u (k^+)^2$, or an extra dependence on $\zeta^2_u$.
The evolution with the renormalization scale $\mu$ is simple:
\begin{equation}
\mu \frac{\partial \phi_+(k^+, k_\perp,\zeta, \mu) }{\partial \mu }
 = (\gamma_q +\gamma_Q) \phi_+(k^+, k_\perp,\zeta, \mu),
\end{equation}
where $\gamma_q$ and $\gamma_Q$ is the anomalous dimension of the light quark field
$q$ and the heavy quark field $h$ in the axial gauge $u\cdot G=0$, respectively.
In the remainder of the paper we will not indicate the $\mu$-dependence explicitly
if it does not cause any confusion.
It should be noted that
one can not simply relate $\phi_+(k^+, k_\perp,\zeta )$ by integrating
$k_\perp$  to the light-cone wave function
in the collinear factorization, whose definition can be found in \cite{LCB}.
The reason for this has been discussed in detail in \cite{MW}.
\par
To perform TMD factorization one needs to calculate the wave function
with perturbative QCD, in which the $B$-meson is replaced by a partonic
state. We take the partonic state $\vert b(m_b v +k_b), \bar
q(k_q)\rangle$ to replace the $B$-meson in the definition,
the momenta are given as
\begin{equation}
k_q^\mu =(k_q^+, k_q^-,\vec k_{q\perp}), \ \ \ \  k_b^\mu =(k_b^+, k_b^-, -\vec k_{q\perp}).
\end{equation}
These partons are on-shell, i.e., $k_q^2 =m_q^2$ and $v\cdot
k_b=0$ in HQET. It should be noted that we take a finite
$k_{q\perp}$ without loosing generality.
The quark mass $m_q$ will regularize collinear singularities. We also introduce
a gluon mass $\lambda$ to regularize infrared singularities.
The variable $k^+$ of the wave function is from
$0$ to $\infty$ in the heavy quark limit.
Actually, from the momentum conservation, it is
from $0$ to $P^+ =m_bv^+ +k_b^+ + k_q^+$. Under the limit $m_b\to
\infty$ we have $P^+ \to \infty$. As discussed in \cite{MW}, if we set $P^+$ to be $\infty$
at the beginning, it may result in some ill-defined distributions.
Therefore we
should take a finite $P^+$ in the calculation and take the limit
$P^+ \to \infty$ in the final result.
For results obtained in this paper we will take the limit where it does not introduce
any problem.
\par
At tree-level, the wave function reads:
\begin{equation}
\phi_+^{(0)} (k^+,k_\perp,\zeta) = \bar v(k_q) \gamma^+ \gamma_5 u(k_b) \delta (k^+- k_q^+)
     \delta^2(\vec k_\perp -\vec k_{q\perp}).
\end{equation}
We will always write a quantity $A$ as $A=A^{(0)} + A^{(1)} +\cdots$, where $A^{(0)}$ and
$A^{(1)}$ stand for tree-level- and one-loop contribution respectively.
At one-loop one can divide the corrections into a real part and a virtual part.
The real part comes from contributions of Feynman diagrams given in Fig.1.
The virtual part comes from contributions of Feynman diagrams given in Fig.2.,
these contributions are proportional to the tree-level result.
\par

\begin{figure}[hbt]
\begin{center}
\includegraphics[width=9cm]{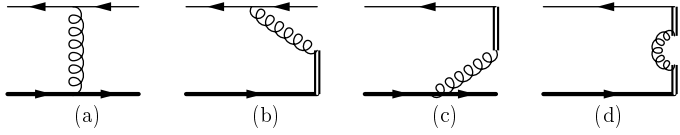}
\end{center}
\caption{Diagrams of one-loop contributions. Thick lines stand for
$b$-quark, double lines represent gauge links.  }
\label{Feynman-dg1}
\end{figure}

\par
To illustrate how to calculate these contributions and how the limit
$u^+<< u^-$ is taken, let us consider the contribution from Fig.1c.
After integrating the $-$-component of the momentum carried by the exchanged
gluon the contribution reads:
\begin{eqnarray}
\phi_+ (k^+,k_\perp, \zeta)\vert_{1c}
  &=&   -\frac{ 2\alpha_s}{3\pi^2}
   \bar v(k_q)  \gamma^+\gamma_5 u(k_b) u\cdot v \theta(k^+-k_q^+)
\nonumber\\
&& \cdot \frac{2 q^+ }{v^+( q_\perp^2 +\lambda^2) +2 v^- (q^+)^2 +i0}
    \cdot \frac{1}{u^+( q_\perp^2 +\lambda^2) +2 u^- (q^+)^2 +i0},
\end{eqnarray}
with $q^+ = k_q^+-k^+$ and $\vec q_\perp = k_{q\perp}-k_{\perp}$.
If we simply set $u^+ =0$, the contribution is proportional
to $1/(k^+ -k_q^+)$ and divergent at $k^+ =k_q^+$. This is the mentioned
light-cone singularity. With the nonzero $u^+$ the contribution is finite
for any $k^+$. The contribution in the limit $u^+ << u^-$ can be derived
by taking the contribution
as a distribution of $k^+$ and it reads:
\begin{eqnarray}
\phi_+ (k^+,k_\perp, \zeta)\vert_{1c}
  &=&  - \frac{ 2\alpha_s}{3\pi^2}
   \bar v(k_q)  \gamma^+\gamma_5 u(k_b)
\nonumber\\
&& \cdot\left [  \left (\frac{\theta(k^+-k_q^+)}{q^+(q_\perp^2 +\lambda^2 + \zeta^2_v (q^+)^2)}
\right )_+
-\delta(k^+ -k_q^+) \frac{1}{2(q_\perp^2 +\lambda^2)} \ln \frac{\zeta_u^2}{ \zeta_v^2 }
\right ] +{\mathcal O }(\zeta_u^{-2}),
\nonumber\\
\zeta_u^2 &=& \frac{2u^-}{u^+} = \frac{\zeta^2}{(k^+)^2},\ \ \ \ \  \zeta_v^2 =\frac{2 v^-}{v^+}.
\end{eqnarray}
In the above the limit $P^+ \to\infty$ is already taken.
From the result we can see that the light-cone singularity is
regularized with the finite but large $\zeta_u^2$. The other contributions of the
real part are:
\begin{eqnarray}
\phi_+ (k^+,k_\perp, \zeta)\vert_{1a} & = &
   \frac{ 2\alpha_s}{3\pi^2}
   \bar v(k_q) \gamma\cdot v (\gamma\cdot(q-k_q) +m_q) \gamma^+\gamma_5 u(k_b)
    F_1
\nonumber\\
   F_1 &= & -i \int\frac{dq^-}{2\pi} \cdot \frac{1}{(q-k_q)^2-m_q^2+i0}\cdot \frac{1}{q^2-\lambda^2+i0}
    \cdot \frac{1}{v\cdot q +i0},
\nonumber\\
\phi_+ (k^+,k_\perp, \zeta)\vert_{1b}
  &=&   \frac{ 2\alpha_s}{3\pi^2}
   \bar v(k_q)  \gamma^+\gamma_5 u(k_b)
\nonumber\\
    && \cdot
     \left [ \frac{k^+}{\Delta_q}
      \left (\frac{\theta (k_q^+ -k^+)}{k_q^+ - k^+}\right )_+
     +\frac{1}{2(q_\perp^2+\lambda^2)} \delta (k^+ -k_q^+)
     \ln \frac{\zeta^2 }{q_\perp^2 +\lambda^2}
     \right ] ,
\nonumber\\
\Delta_q &=& k_q^+ ( (q_\perp -xk_{q\perp})^2 +x^2 m_q^2 +(1-x)\lambda^2), \ \ \  x = 1- \frac{k^+}{k_q^+},
\nonumber\\
\phi_+ (k^+,k_\perp, \zeta)\vert_{1d} &=& -\frac{ 2\alpha_s}{3\pi^2}
   \bar v(k_q) \gamma^+\gamma_5 u(k_b)\delta (k^+-k_q^+)
    \frac{ 1}{q^2_\perp +\lambda^2  }.
\end{eqnarray}
The integral $F_1$ for the contribution from Fig.1a can be done easily,
but it results in a lengthy expression. We will show later that the
contribution will not affect the perturbative coefficient function $H$.
\par

\begin{figure}[hbt]
\begin{center}
\includegraphics[width=8cm]{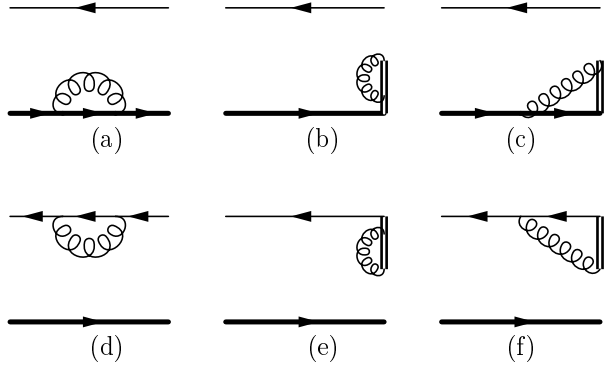}
\end{center}
\caption{The virtual part of the one-loop correction. }
\label{Feynman-dg2}
\end{figure}

\par
The virtual part of the one-loop correction is from the Feynman diagrams
given in Fig.2. Contributions from each diagrams are:
\begin{eqnarray}
\phi_+ (k^+,k_\perp, \zeta)\vert_{2a} &=& \phi_+ (k^+,k_\perp, \zeta)\vert_{2b} =
    \phi_+ (k^+,k_\perp, \zeta)\vert_{2e} = \phi_+^{(0)} (k^+,k_\perp, \zeta)
     \cdot \frac{\alpha_s}{3\pi}
    \ln \frac{\mu^2}{\lambda^2},
\nonumber\\
\phi_+ (k^+,k_\perp, \zeta)\vert_{2c} &=& -\phi_+^{(0)} (k^+,k_\perp, \zeta) \cdot \frac{\alpha_s}{3\pi}
    \ln \frac{\mu^2}{\lambda^2}\ln\frac{\zeta_u^2}{\zeta^2_v},
\nonumber\\
\phi_+ (k^+,k_\perp, \zeta)\vert_{2f} &=& \phi_+^{(0)} (k^+,k_\perp, \zeta)
 \cdot \frac{\alpha_s}{6\pi}
    \left [ 2\ln\frac{\mu^2}{m_q^2} + 2\ln\frac{\zeta^2}{m_q^2}
       -\ln^2 \frac{\zeta^2}{m_q^2}
       -2\ln\frac{m_q^2}{\lambda^2}\ln\frac{\zeta^2}{m_q^2}
     -\frac{2\pi^2}{3} +4 \right ],
\nonumber\\
\phi_+ (k^+,k_\perp, \zeta)\vert_{2d} &=& \phi_+^{(0)} (k^+,k_\perp, \zeta) \cdot \frac{\alpha_s}{6\pi}
    \left [ -\ln \frac{\mu^2}{m_q^2} +2 \ln\frac{m_q^2}{\lambda^2} -4 \right ] ,
\end{eqnarray}
The complete one-loop contribution $\phi_+^{(1)}$ is the sum of contributions from the 10 Feynman
diagrams in Fig.1. and Fig.2. With these results one can derive the evolution
of $\zeta$. For this we transform the wave-function into the impact parameter $b$-space:
\begin{equation}
\phi_+(k^+, b, \zeta, \mu) = \int d^2 k_\perp e^{i \vec k_\perp \cdot \vec b}
       \phi_+(k^+,k_\perp, \zeta, \mu),
\end{equation}
the evolution reads:
\begin{eqnarray}
\zeta \frac{\partial}{\partial \zeta} \phi_+(k^+,b, \zeta,\mu) =
   \left [-\frac{4\alpha_s}{3\pi} \ln\frac{\zeta^2 b^2 e^{2\gamma-1}}{4}
       -\frac{2\alpha_s}{3\pi} \ln\frac{\mu^2e}{\zeta^2} \right ]
       \phi_+(k^+,b,\zeta,\mu).
\end{eqnarray}
The first factor is the famous factor $K+G$\cite{CS,CSS},
the last factor comes because we used HQET for the heavy quark.
\par
Before ending the section, we briefly discuss the heavy quark limit $P^+\to\infty$.
For the usual light-cone wave function, this limit will result in that
the wave function is not normalizable as found in \cite{LCB,LN} and it is shown
through an explicit calculation with perturbative theory in \cite{MW}.
For the TMD light-cone wave function it is normalizable if the transverse
momentum is not integrated out. When we transform the TMD light-cone wave function
into $b$-space, we should keep $P^+$ finite.
\par\vskip20pt
\noindent
{\bf\large 3. Notations and Factorization at Tree Level}
\par\vskip15pt
We consider the radiative decay of the $B$-meson $\bar B$ which contains at least a $b$-quark
and a light anti-quark $\bar q$:
\begin{equation}
   \bar B \to \gamma + \ell +\bar \nu.
\end{equation}
We take a frame in which $\bar B$ moves in the $z$-direction with
the velocity $v^\mu =(v^+, v^-,0,0)$ and the photon with the momentum
$p^\mu = (0,p^-,0,0)$. It is worth to mention here that
this decay has not been observed so far. An upper bound for the branching ratio is given in \cite{Bexp}:
\begin{equation}
 {\rm Br} (\bar B \to \gamma + \ell +\bar \nu ) < 2.0\times 10^{-6}.
\end{equation}
\par
In the decay the effect of the strong interaction
is controlled by a matrix element of the operator
$\bar q \gamma^\mu (1-\gamma_5) b$
with $b(x)$ being the $b$-quark field in the full QCD.
Since we use HQET for the heavy $b$-quark, the matrix element
can be matched to HQET:
\begin{equation}
\langle \gamma(\epsilon^*, p) \vert \bar q  \gamma^\mu (1-\gamma_5) b \vert \bar B(v)\rangle
= f(\mu) \langle \gamma(\epsilon^*, p) \vert \bar q  \gamma^\mu (1-\gamma_5) h \vert \bar B(v)\rangle,
\end{equation}
where $f(\mu)$ is the matching coefficient. It is
given by:
\begin{equation}
   f(\mu) = 1 + \frac{\alpha_s(\mu)}{3\pi} \left ( 3 \ln\frac{m_b}{\mu} -2 \right ) + {\mathcal O}(\alpha_s^2).
\end{equation}
The HQET matrix element
can be parameterized as
\begin{eqnarray}
{\mathcal T}^\mu &=& \frac{1}{\sqrt{4\pi\alpha}}
\langle \gamma(\epsilon^*, p) \vert \bar q  \gamma^\mu (1-\gamma_5) h \vert \bar B(v)\rangle
\nonumber\\
  &=& \varepsilon^{\mu\nu\rho\sigma} \epsilon^*_\nu v_\rho p_\sigma F_V (v\cdot p)
    + i \left (v\cdot p \epsilon^*_\mu - v\cdot \epsilon^* p^\mu \right ) F_A(v\cdot p).
\end{eqnarray}
In the rest frame of $\bar B$, $v\cdot p$ is the energy of the photon. The invariant
$v\cdot p$ can be from 0 to $M_B/2$. The photon is emitted by quarks inside the $B$-meson.
If $v\cdot p$ is large, i.e., $v\cdot p \gg \Lambda_{QCD}$ those quarks will change their momenta significantly, i.e.,
the emission becomes a short-distance process. This leads to that
those form factors, hence the matrix element can be studied with perturbative QCD, in which
one can separate short-distance- and long distance effect by factorization.
\par

\begin{figure}[hbt]
\begin{center}
\includegraphics[width=6cm]{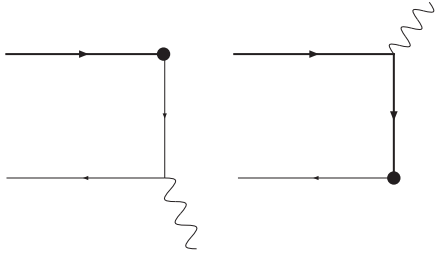}
\end{center}
\caption{Tree-level contribution to the matrix element. The thick line is for the $b$-quark, the black
dot denotes the insertion of the operator.  }
\label{Feynman-dg3}
\end{figure}
\par
\par
To show the factorization, one usually replaces hadronic states with reasonable parton states,
then calculate processes which need to be factorized and nonperturbative objects like
wave functions in our case to extract the perturbative coefficient functions. A factorization
means at least that those coefficient functions do not contain any soft divergence.
For our purpose we replace the $\bar B$ state $\vert \bar B \rangle$
with the partonic state $\vert \bar q b\rangle$. The momenta of the partons are the same
as in Eq.(5). At tree-level the contribution to the matrix element
is given by the two diagrams in Fig.3. The second diagram will not contribute in the
heavy quark limit by noting the fact $v\cdot \epsilon^* =0$ for a real photon.
The tree-level amplitude $T^\mu$ reads:
\begin{equation}
  T_\mu^{(0)} = Q_{q} \bar v(k_q) \gamma\cdot \epsilon^* \cdot
   \frac { \gamma \cdot (p-k_q) +m_q}{(p-k_q)^2 -m_q^2 } \gamma_\mu (1-\gamma_5) u(k_b),
\end{equation}
where $Q_{q}$ is the charge fraction of $\bar q$.
In TMD factorization one will neglect the transverse momentum of initial partons in nominators of propagators
but keep
it in the denominators. The case studied here is rather special because the denominator does not
depend on the transverse momentum. With a little algebra one can show that
\begin{eqnarray}
T_\mu^{(0)} &=& -\frac{Q_q}{2p\cdot k_q} \bar v(k_q) \gamma\cdot \epsilon^* \gamma\cdot p \gamma_\mu (1-\gamma_5) u(k_b)
                  +\cdots
\nonumber\\
  &=& \frac{i Q_q}{2 v\cdot p} \left [
 \varepsilon_{\mu\nu\rho\sigma} \epsilon^{*\nu} v^\rho p^\sigma
    + i \left (v\cdot p \epsilon^*_\mu - v\cdot \epsilon^* p_\mu \right ) \right ] \cdot \frac{1} {k_q^+}
  \bar v(k_q) \gamma^+ \gamma_5 u(k_b) + \cdots,
\end{eqnarray}
where $\cdots$ denotes the neglected $k_{q\perp}$-dependence from the quark propagator
and the contribution from the partonic state which does not have the same quantum numbers
as $\bar B$ does.
\par
With the tree-level result of the TMD light-cone wave function, we obtain
the factorization for those form factors:
\begin{equation}
 F_V = F_A =\frac{i Q_q}{2 v\cdot p} \int dk^+ d^2 k_\perp \phi_+ (k^+,k_\perp,\zeta,\mu) \frac{1}{k^+}.
\end{equation}
At the orders considered in the work, $F_V$ is always the same as $F_A$.
We will write our factorization formulas as:
\begin{equation}
F_V = F_A =\frac{i Q_q}{2 v\cdot p}  \int dk^+ d^2 k_\perp d l^+ d ^2 l_\perp \phi_+(k^+,k_\perp, \zeta)
         \tilde S (l^+, l_\perp, \zeta_u) \theta (k^+ +l^+) H (k^+ +l^+, \zeta_u),
\end{equation}
so that at the leading order of $\alpha_s$  the perturbative coefficient function $H$
and the soft factor in perturbation theory
at tree-level reads:
\begin{equation}
  H^{(0)}(k^+,\zeta_u) = \frac{1}{k^+}, \ \ \ \ \  \tilde S^{(0)}(k^+, k_\perp,\zeta_u ) =\delta (k^+) \delta^2(\vec k_\perp).
\end{equation}
It is noted that at the leading order $H$ does not depend on $k_\perp$ in the case studied here,
while in the other cases like $B\to \pi$ transition it does. If one replaces the B-meson state with
a partonic state of off-shell partons, one can have a $H^{(0)}$ which depends on $k_\perp$\cite{NaLi}.
But the amplitude $T^\mu$ with the state of off-shell partons is not gauge-invariant.
In general it is not clear if the factorization with such a state can be made in a gauge invariant way.
\par
At tree-level one can not
determine the form of the soft factor $\tilde S$,
because it is designed to subtract infrared divergences at higher
orders of $\alpha_s$. It should be a $\delta$-function at tree level.
At one-loop level with the partonic state, the factorization formula takes the form:
\begin{equation}
T^{(1)}_\mu \sim \phi_+^{(0)} \otimes \tilde S^{(0)}  \otimes H^{(1)}
              +\phi_+^{(0)} \otimes \tilde S^{(1)}  \otimes H^{(0)}
              +\phi_+^{(1)} \otimes \tilde S^{(0)}  \otimes H^{(0)},
\end{equation}
the soft factor should be chosen so that all soft divergences of $T^{(1)}_\mu$ are contained in
the second- and third term and $H^{(1)}$ is free from any soft divergence.
The soft factor should also be chosen so that the factorization can be extended beyond one-loop level.

\par\vskip20pt
\noindent
{\bf\large 4. The Soft Factor and Factorization at One-Loop Level}
\par\vskip15pt
In this section we will perform TMD factorization at one-loop level and determine the operator form of the
soft factor. The perturbative coefficient function will also be determined at one-loop level.
Let us first consider the one-loop corrections to the amplitude ${\mathcal T}^\mu$.
The corrections are from diagrams given in Fig.4.
\par\vskip20pt

\begin{figure}[hbt]
\begin{center}
\includegraphics[width=13cm]{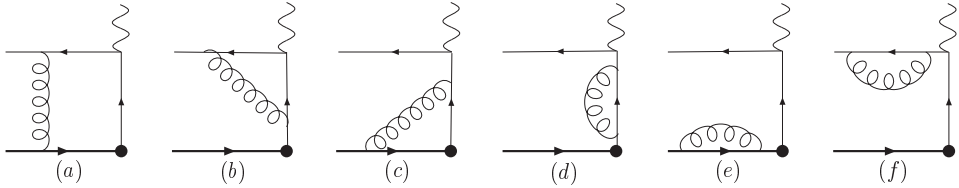}
\end{center}
\caption{One-loop contribution to the matrix element. The thick line is for the $b$-quark, the black
dot denotes the insertion of the operator.  }
\label{Feynman-dg4}
\end{figure}
\par\vskip20pt
The contribution from Fig.4a reads:
\begin{eqnarray}
T^{\mu} |_{4a} &=& iQ_q g_s^2 C_F \int \frac{d^4 l}{(2\pi)^4} \bar{v}(k_q)
\gamma \cdot v \frac{\gamma \cdot (k_q +l)}{(k_q +l)^2 -m_q^2 +i\varepsilon}
 \gamma \cdot \epsilon^* \frac{\gamma \cdot (p-k_q -l)}{(p-k_q-l)^2 -m_q^2
+i\varepsilon} \nonumber \\
&& \hspace{5mm} \cdot \gamma^{\mu} (1-\gamma_5) \frac{1}{-v \cdot l
+i\varepsilon}
\frac{1}{l^2 -\lambda^2 +i\varepsilon} u(k_b) \nonumber \\
&=& iQ_q g_s^2 C_F \int \frac{d^4 l}{(2\pi)^4} \bar{v}(k_q) \gamma \cdot v
\frac{\gamma \cdot (k_q +l)}{(k_q +l)^2 -m_q^2 +i\varepsilon} \gamma \cdot
\epsilon^* \gamma \cdot p \gamma^{\mu} (1-\gamma_5) u(k_b) \nonumber \\
&& \hspace{5mm} \cdot \frac{1}{-2p \cdot (k_q +l) +i\varepsilon} \cdot
\frac{1}{-v \cdot l +i\varepsilon} \cdot
\frac{1}{l^2 -\lambda^2 +i\varepsilon} + {\cal O} (E_{\gamma}^{-2} ).
\end{eqnarray}
It is easy to find that this contribution up to a power correction is exactly represented by the contribution
from $\phi_+\vert_{1a}$ to the third term in Eq.(23). Therefore, this contribution and $\phi_+\vert_{1a}$
is irrelevant for the determination of $H^{(1)}$ and $\tilde S^{(1)}$.
Also, the contributions from Fig.4e and Fig.4f to $T^{(1)}_\mu$ are reproduced
by the contributions from $\phi_+\vert_{2a}$ and $\phi_+\vert_{2d}$ in the third term in Eq.(23), respectively.
The contributions from other diagrams to $T^{(1)}_\mu$ are:
\begin{eqnarray}
T^\mu\vert_{4b} &=&  T_\mu^{(0)}\cdot \frac{\alpha_s}{3\pi} \left [
       \ln \frac{\mu^2}{2 k_q\cdot p} + 2 \ln \frac{2k_q\cdot p}{m_q^2}  \right ],
\nonumber\\
T_\mu\vert_{4c} &=& T_\mu^{(0)} \cdot \frac{\alpha_s}{3\pi}
 \left [ -\ln^2\left(\frac{2 p\cdot k_q}{\zeta_v^2 (k_q^+)^2}\right )
           + \ln\frac{\mu^2}{\zeta_v^2 (k_q^+)^2} +2 -\frac{4}{3} \pi^2 \right ],
\nonumber\\
T^\mu\vert_{4d} &=&T_\mu^{(0)}\cdot \frac{\alpha_s}{3\pi} \left [
      \ln \frac{2k_q\cdot p}{\mu^2} -1 \right ].
\end{eqnarray}
Our $T^\mu\vert_{4b}$ agrees with that of \cite{DS}, but is not in agreement with that in \cite{KPY}.
The other contributions are in agreement with \cite{KPY}.
In these contributions there are no infrared singularities. They have only a collinear singularity
from Fig.4b, represented by $\ln m_q$. The relevant contributions to
$\phi_+^{(1)} \otimes \tilde S^{(0)} \otimes H^{(0)}$
by using $\tilde S^{(0)}$  are:
\begin{eqnarray}
W_{b} &=& \int dk^+ d^2 k_\perp \frac{1}{k^+} \left [
   \phi_+ \vert_{1b} (k^+,k_\perp) + \phi_+ \vert_{2e} (k^+,k_\perp) +\phi_+ \vert_{2f} (k^+,k_\perp)
\right ] / \bar v(k_q)\gamma^+\gamma_5 u(k_b)
\nonumber\\
 &=& \frac{\alpha_s}{3\pi k_q^+} \left \{ 2 + \ln \frac{\mu^2}{\zeta^2}
      -\frac{1}{2} \ln^2 \frac{\lambda^2}{\zeta^2} +2 \ln\frac{\zeta^2}{m_q^2} +\ln \frac{\mu^2}{\lambda^2}
      -\frac{1}{2} \pi^2  -
            \int \frac{d^2 k_\perp}{\pi} \frac{1}{\lambda^2 + k_\perp^2} \ln\frac{\lambda^2 + k_\perp^2}{\zeta^2}\right\},
\nonumber\\
W_{c} &=& \int dk^+ d^2 k_\perp \frac{1}{k^+} \left [
   \phi_+ \vert_{1c} (k^+,k_\perp) + \phi_+ \vert_{2b} (k^+,k_\perp) +\phi_+ \vert_{2c} (k^+,k_\perp)
\right ]/ \bar v(k_q)\gamma^+\gamma_5 u(k_b)
\nonumber\\
&=&\frac{\alpha_s}{3\pi k_q^+} \left \{ \ln\frac{\mu^2}{\lambda^2}
+\ln\frac{\mu^2}{\lambda^2}\ln \frac{\zeta_v^2}{\zeta_u^2}
-\frac{5}{6} \pi^2 -\frac{1}{2} \ln^2 \frac{\lambda^2}{\zeta_v^2(k_q^+)^2}
-\int \frac{d^2 k_\perp}{\pi} \frac{1}{\lambda^2 + k_\perp^2} \ln\frac{\lambda^2 + k_\perp^2}{\zeta^2}\right\},
\nonumber\\
W_{d} &=&\int dk^+ d^2 k_\perp \frac{1}{k^+}\phi_+ \vert_{1d}(k^+,k_\perp)/ \bar v(k_q)\gamma^+\gamma_5 u(k_b)
 =-\frac{2\alpha_s}{3\pi k_q^+}\int \frac{d^2 k_\perp}{\pi} \frac{1}{\lambda^2 + k_\perp^2}.
\end{eqnarray}
Comparing the above two equations, we find that the collinear singularity from Fig.4b. is reproduced
by the contribution in $W_b$ from Fig.2f. But, there are many infrared singularities
in $\phi_+^{(1)} \otimes \tilde S^{(0)} \otimes H^{(0)} \sim W_a+ W_b + W_c + W_d +W_e$, where
$W_a$ is the contribution from Fig.1a and $W_e$ is the sum of contributions from Fig.2a and Fig.2d.
There are even
ultraviolet divergences. However these divergences are closely related to corresponding infrared singularities,
as they stand. Once these infrared singularities are subtracted, one can expect that those ultraviolet divergences
are subtracted too. As mentioned before, the contributions of $W_a$ and $W_e$ represent
those to the $T^\mu$ from Fig.4a, Fig.4e and Fig.4f. To complete the factorization
one needs to find the soft factor so that all infrared singularities in $W_b+W_c+W_d$ and also the divergent integrals
over $k_\perp$ are subtracted by the soft factor.
\par
Clearly all these infrared singularities are from the TMD wave functions, i.e., from contributions
from Fig.1. and Fig.2. By using the eikonal approximation one easily finds that these singularities can be
reproduced by the expectation value of the product of gauge links:
\begin{eqnarray}
  S_4 (q^+, q_\perp ) &=& \int d z^- d^2 z_\perp e^{i q^+ z^- - i \vec q_\perp \cdot \vec z_\perp }
            S_4 (z^-, z_\perp),
\nonumber\\
  S_4 (z^-, z_\perp) &=&
                \frac{1}{3} {\rm Tr} \langle 0 \vert T \left [ L_{\tilde u}^\dagger (z,-\infty)
                 L_{u}^\dagger (\infty,z)L_{u} (\infty,0) L_{v} (0,-\infty) \right ] \vert 0\rangle \vert_{z^+ =0},
\nonumber\\
 L_{\tilde u}(z,-\infty) &=& P \exp \left ( -i g_s \int^{0}_{-\infty} d\lambda
     \tilde u\cdot G (\lambda \tilde u+ z ) \right ),
\nonumber\\
 L_{v}(z,-\infty) &=& P \exp \left ( -i g_s \int^{0}_{-\infty} d\lambda
     v\cdot G (\lambda v+ z ) \right ),
\end{eqnarray}
where the direction in $L_{\tilde u}$ is chosen as $\tilde u^+ >>\tilde u^-$. The gauge
link $ L^\dagger_{\tilde u} $ just simulates an anti-quark $\bar q$ in the initial state
and $L_{v}$ the $b$-quark in  the initial state. If one only takes this product
of the gauge links into account, one can expect that the quantity
\begin{equation}
 \frac{ \phi_+(z^-, b,\zeta,\mu)}{ S_4 (z^-, b)}
\end{equation}
is free from infrared singularities. This is checked at one-loop level.
However, because part of contributions
from $\phi_+^{(1)}$,  which are from Fig.1a, Fig.2a and Fig.2d, is already used up to subtract
soft divergences in $T_\mu^{(1)}$, as discussed before, one can not expect that the soft factor
$\tilde S$ can be formed with $S_4$ only. We will turn to this point later and concentrate
at moment on the perturbative results for $S_4$.
\par
At tree-level, the result is just a $\delta$-function:
\begin{equation}
  S_4^{(0)} (q ^+, q_\perp ) =\delta (q^+) \delta^2 (\vec q_\perp).
\end{equation}
At one-loop level, there are contributions from diagrams which have a one-to-one correspondence
to those diagrams given in Fig.1. and Fig.2., in which one only needs to replace
the light-quark line with the double line of the gauge link $ L^\dagger_{\tilde u} $.
The corresponding contributions as a distribution of $q^+$ for the range
$-k^+ < q^+ <\infty$ under the limits $u^+ \to 0$ and $\tilde u^- \to 0$ are:
\begin{eqnarray}
S_4(q^+,\vec q_\perp) {\large \vert}_{1a} &=& \frac{2\alpha_s}{3 \pi^2} \left [
    \frac{ \theta (q^+)}{q^+( q^2_\perp +\lambda^2+\zeta_{v}^2 (q^+)^2)}
     + \frac{\theta (-q^+)} {q^+ (q^2_\perp +\lambda^2 +\zeta^2_{\tilde u} (q^+)^2 )} \right ] +"{\rm imaginary\ part}",
\nonumber\\
S_4(q^+,\vec q_\perp) {\large \vert}_{1b} &=& -\frac{2\alpha_s}{3 \pi^2}
  \left[ \frac{ 1}{q^2_\perp +\lambda^2+\zeta^2_{\tilde u} (q^+)^2} \theta(-q^+) \left ( \frac{1}{q^+} \right )_+
      - \frac{1}{2}\delta(q^+) \frac{ 1}{q^2_\perp +\lambda^2}\ln \frac{\zeta^2}{q^2_\perp +\lambda^2} \right ]
\nonumber\\
S_4(q^+,\vec q_\perp) {\large \vert}_{1c} &=& \frac{2\alpha_s}{3 \pi^2}
 \left [ \left (\frac{ \theta (q^+) }{ q^+ (q^2_\perp +\lambda^2 +\zeta_v^2 (q^+)^2) } \right )_+
    + \frac{1}{2}\delta (q^+) \frac{1}{q^2_\perp +\lambda^2} \ln \frac{\zeta^2_u}{\zeta_v^2}
    \right ],
\nonumber\\
S_4(q^+,\vec q_\perp) {\large \vert}_{1d} &=& -\frac{2\alpha_s}{3 \pi^2}
              \frac{\delta (q^+) }{q^2_\perp +\lambda^2 },
\end{eqnarray}
and the contributions from the diagrams corresponding to those in Fig.2 are:
\begin{eqnarray}
S_4(q^+,\vec q_\perp)\vert_{2a} &=&S_4(q^+,\vec q_\perp)\vert_{2b} = S_4(q^+,\vec q_\perp)\vert_{2d}=
         S_4(q^+,\vec q_\perp)\vert_{2e}
        = S_4^{(0)}(q^+,\vec q_\perp) \cdot \frac{\alpha_s}{3\pi}
    \ln \frac{\mu^2}{\lambda^2},
\nonumber\\
S_4(q^+,\vec q_\perp)\vert_{2c} &=& -S_4^{(0)}(q^+,\vec q_\perp) \frac{\alpha_s}{3\pi}
    \ln \frac{\mu^2}{\lambda^2}\ln\frac{\zeta_{u}^2}{\zeta^2_v},
\nonumber\\
S_4(q^+,\vec q_\perp)\vert_{2f} &=& -S_4^{(0)}(q^+,\vec q_\perp) \frac{\alpha_s}{3\pi}
    \ln \frac{\mu^2}{\lambda^2}\ln\frac{\zeta_u^2}{\zeta^2_{\tilde u}},
\end{eqnarray}
with $\zeta^2_{\tilde u} = 2\tilde u^-/\tilde u^+$. If we identify the soft factor $\tilde S(z^-, \vec b)$ as
$S_4^{-1} (z^-,\vec b)$, their contributions to $\phi_+^{(0)}\otimes \tilde S^{(1)}\otimes H^{(0)}$ can be grouped
similarly as those to $\phi_+^{(1)}\otimes \tilde S^{(0)}\otimes H^{(0)}$. They are:
\begin{eqnarray}
U_a &=&
-\int dl^+  d^2 l_\perp \frac{1}{k_q^+ +l^+}
S_4\vert_{1a}(l^+,l_\perp)
\nonumber\\
&=& \frac{\alpha_s}{3\pi k_q^+}
\left\{ \frac{1}{2} \mbox{ln}^2 \frac{\zeta_v^2(k_q^+)^2}{\lambda^2}
+\frac{5\pi^2}{6} -\frac{1}{2} \mbox{ln}^2
\frac{\zeta_{\tilde u}^2(k_q^+)^2}{\lambda^2}
 + \Delta  \right \} +"{\rm imaginary\ part}",
\nonumber\\
U_b &=&
-\int dl^+ d^2 l_\perp \frac{1}{k_q^+ +l^+} \left [ S_4\vert_{1b} + S_4\vert_{2e}+ S_4\vert_{2f} \right ](l^+,l_\perp)
\nonumber\\
 &=& \frac{\alpha_s}{3\pi k_q^+} \left \{ -\Delta +\frac{1}{2} \ln^2\frac{\lambda^2}{\zeta^2_{\tilde u} (k_q^+)^2}
          -\ln\frac{\mu^2}{\lambda^2}- \ln\frac{\mu^2}{\lambda^2} \ln\frac{\zeta^2_{\tilde u}}{\zeta^2_u}
                  +\int \frac{d^2 k_\perp}{\pi} \frac{1}{k_\perp^2 +\lambda^2} \ln\frac{k_\perp^2+\lambda^2}{\zeta^2}
          \right \},
\nonumber\\
U_c &=& -\int dl^+  d^2 l_\perp \frac{1}{k_q^+ +l^+}
 \left [ S_4\vert_{1c} + S_4\vert_{2b}+ S_4\vert_{2c} \right ](l^+,l_\perp)
  \nonumber\\
  &=& \frac{\alpha_s}{3\pi k_q^+} \left \{
     \frac{5}{6}\pi^2 +\frac{1}{2} \ln^2\frac{\lambda^2}{\zeta^2_v (k_q^+)^2}
   -\ln\frac{\mu^2}{\lambda^2}-\ln\frac{\mu^2}{\lambda^2} \ln\frac{\zeta^2_v}{\zeta^2_u}
         + \int\frac{d^2 k_\perp}{\pi} \frac{1}{k_\perp^2 +\lambda^2} \ln\frac{k_\perp^2+\lambda^2}{\zeta^2} \right \},
\nonumber\\
U_d &=&-\int dl^+  d^2 l_\perp \frac{1}{k_q^+ +l^+}
S_4\vert_{1d}(l^+,l_\perp)
  =\frac{2\alpha_s}{3\pi k_q^+} \int \frac{d^2 k_\perp}{\pi}  \frac{1}{k_\perp^2 +\lambda^2} ,
\nonumber\\
U_e &=& -\int dl^+ d^2 l_\perp \frac{1}{k_q^+ +l^+} \left [ S_4\vert_{2a} + S_4\vert_{2d} \right ](l^+,l_\perp)
= -\frac{2\alpha_s}{3\pi} \ln\frac{\mu^2}{\lambda^2},
\end{eqnarray}
where $\Delta$ is a divergent quantity:
\begin{equation}
  \Delta = \lim_{k^+ \to 0} \ln\frac{k_q^+}{k^+}
  \int \frac{d^2 k_\perp}{\pi} \frac{2}{\zeta^2_{\tilde u} (k_q^+)^2 + k_\perp^2},
\end{equation}
which will be cancelled in $U_a+U_b$. Comparing the sum  $U_a+U_b+U_c+U_d+U_e$ with
$W_b+W_c +W_d$, we note first that the divergent integrals over $k_\perp$
in $W_b$, $W_c$ and $W_d$ are completely subtracted by those in $U_b$, $U_c$ and $U_d$,
respectively. Also the infrared singularities with $\ln \lambda$ in
$W_b$, $W_c$ and $W_d$ are completely subtracted by those in $U_b$, $U_c$ and $U_d$, respectively.
The remaining infrared singularities are only from $U_a$ and $U_e$.
\par


\begin{figure}[hbt]
\begin{center}
\includegraphics[width=8cm]{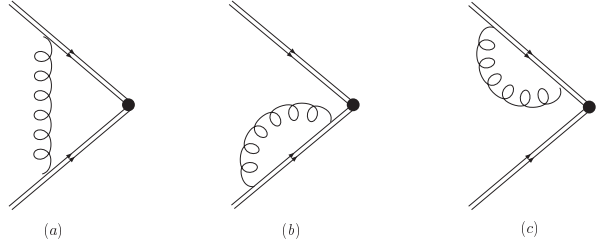}
\end{center}
\caption{One-loop contribution to $S_2$. The double lines represent the two gauge links. One is for
$L_v$, the other one is for $L^\dagger_{\tilde u}$.  }
\label{Feynman-dg5}
\end{figure}
\par\vskip20pt
These remaining singularities can be reproduced by the product of the gauge links:
\begin{equation}
 S_2 =\frac{1}{3} {\rm Tr} \langle 0 \vert T \left [ L^\dagger_{\tilde u}(0,-\infty) L_v(0,-\infty) \right ] \vert 0 \rangle.
\end{equation}
At leading order $S_2^{(0)}=1$. At one-loop level, the contributions are from the diagrams given in Fig.5.
\begin{eqnarray}
  S_2\vert_{5a} &=& -\frac{\alpha_s}{3\pi}\ln \frac{\mu^2}{\lambda^2} \ln \frac{\zeta^2_v}{\zeta^2_{\tilde u}}
               +"{\rm imaginary\ part}",
\nonumber\\
S_2 \vert_{5b} &=& S_2\vert_{5c} = \frac{\alpha_s}{3\pi} \ln \frac{\mu^2}{\lambda^2}.
\end{eqnarray}
Now we turn to the imaginary or absorptive part. In the amplitude $T^\mu$ it has an absorptive part
from the box diagram Fig.4a and its contribution is already contained
in the contribution from the TMD wave function in Fig.1a. The remaining parts $T^\mu$ can not
have an absorptive part. Therefore, one should eliminate possible absorptive part
in the soft factor. At one-loop level, the imaginary part from $S_4$ is the same
as that from $S_2$. But this is from perturbative theory. To eliminate the absorptive
part one can simply take the real parts of those products of gauge links.
Therefore we determine the soft factor as:
\begin{eqnarray}
\tilde S(z^-, \vec b, \zeta_u, \mu  )  &=&  \frac{ {\rm Re}\left [ S_2 (\zeta_{\tilde u}, \mu)\right ]}
      {{\rm Re}\left [ S_4 (z^-, \vec b, \zeta_u,\zeta_{\tilde u}, \mu)\right ]},
\nonumber\\
\tilde S (k^+, k_\perp, \zeta,\mu) &=& \frac{1}{(2\pi)^3}\int d z^- d^2 b e^{ik^+ z^- -i \vec k_\perp \cdot
             \vec b }\tilde S(z^-, \vec b, \zeta_u, \mu  ).
\end{eqnarray}
It should be noted that $S_2$ and $S_4$ depend on $\zeta_{\tilde u}$, but the soft factor
as the ratio of them does not depend on $\zeta_{\tilde u}$.
With the defined soft factor, the form factors can be factorized as in Eq.(21). They take a
compact form in the $b$-space:
\begin{equation}
F_V = F_A =\frac{i Q_q}{2 v\cdot p} \lim_{b\to 0} \int dk^+  d l^+  \phi_+(k^+,b, \zeta, \mu)
         \tilde S (l^+, b, \zeta_u, \mu) \theta (k^+ +l^+) H (k^+ +l^+, \zeta_u, \mu).
\end{equation}
The limit $b\to 0$ should be taken after the integrations.
With the results presented before, we determine the perturbative coefficient function $H$ as:
\begin{eqnarray}
\label{H1}
H(k^+, \zeta_u,\mu) &=& \frac{1}{k^+} \left \{ 1 +  \frac{2\alpha_s(\mu) }{3\pi} \left [ -\frac{1}{2} \mbox{ln}^2
\frac{2 k \cdot p}{\zeta_v^2 (k^+)^2} +\frac{1}{4} \mbox{ln}
\frac{\zeta_u^2}{\zeta_v^2} \left( \mbox{ln} \frac{\zeta_u^2 (k^+)^2}{\mu^2}
+ \mbox{ln} \frac{\zeta_v^2 (k^+)^2}{\mu^2} \right )
\right. \right. \nonumber \\
&& \left. \left.
+\frac{1}{2} \mbox{ln} \frac{2 k \cdot p}{\zeta_v^2 (k^+)^2}
 +\frac{1}{2} \mbox{ln} \frac{2 k \cdot p}
{\zeta_u^2 (k^+)^2}  -\frac{1}{2}-\frac{5\pi^2}{6} \right ] \right\} +{\mathcal O} (\alpha_s^2),
\end{eqnarray}
which is free from any soft divergence.
All soft singularities are cancelled
on a diagram-by-diagram basis. The cancellation on a diagram-by-diagram basis is important
for extending the factorization beyond one-loop level. General arguments for the factorization
at any loop can be given by performing an analysis of relevant reduced diagrams and infrared power-counting.
The perturbative coefficient function $H$ here does not contain the double log $\ln^2 \mu^2$ in contrast
with that in the collinear factorization\cite{DS,LPW,BHLN}, instead of $\ln^2 \mu^2$  it contains
$\ln^2\zeta_u^2$ and other log terms. All of those log terms need to be resummed if they can be large.
\par
It should be noted that for the case studied here one may redefine the TMD light-cone wave function
by including the soft factor as $\phi_+' = \phi_+ \otimes \tilde S$, so that the form factors
take the form $\phi_+' \otimes H$. Then our results look similar to those in the collinear factorization.
However, it is not clear if the same can be done for other processes, because they have not been studied yet.
Therefore, we leave the soft factor there explicitly.
\par\vskip20pt
\noindent
{\bf\large 5. Relation between Two Factorizations}
\par\vskip15pt
In the section we show that the result of the collinear factorization for the decay
can be obtained from that of TMD factorization, which is given in the last section.
Hence, a simple relation between two factorizations is found for the decay.
\par
In collinear factorization, the transverse momenta of partons are integrated
out and the collinear light-cone wave function can be defined as\cite{LCB}:
\begin{equation}
\Phi_+(k^+,\mu) =\int \frac{d z^- }{2\pi}  e^{ik^+z^-}
\langle 0 \vert \bar q(z^- n) L_n^\dagger (\infty, z^-n)
  \gamma^+ \gamma_5 L_n (\infty,0) h(0) \vert \bar B(v) \rangle,
\end{equation}
with the gauge link $L_n$ defined with the light-cone vector $n^\mu =(0,1,0,0)$:
\begin{equation}
L_n (\infty, z) = P \exp \left ( -i g_s \int_{0} ^{\infty} d\lambda
     n\cdot G (\lambda n + z ) \right ) .
\end{equation}
By taking the same partonic state as given in Sect.2., the wave function can be calculated
with perturbative QCD. The result can be found in \cite{MW}. With this result
and that in the last section, one can easily derive the result in the collinear factorization:
\begin{equation}
F_V = F_A =\frac{i Q_q}{2 v\cdot p}  \int dk^+   \Phi_+(k^+, \mu)
          H_c (k^+, \mu),
\end{equation}
where $H_c$ is the perturbative coefficient function and is given by:
\begin{eqnarray}
H_c(k^+, \mu) &=& \frac{1}{k^+} \left \{
  1 +\frac{\alpha_s}{3\pi} \left [
    \frac{1}{2}\ln^2\frac{\mu^2}{\zeta_v^2 (k^+)^2} +  \ln\frac{2 k\cdot p}{\mu^2}
    +\ln\frac{2 k\cdot p}{\zeta_v^2 (k^+)^2} -  \ln^2 \frac{2k\cdot p}{\zeta_v^2 (k^+)^2}
\right.\right.
\nonumber\\
   && \left.\left.
      -3 -\frac{7\pi^2}{12} \right ] \right \} + {\mathcal O} (\alpha_s^2),
\end{eqnarray}
where the logarithmic terms agree with those in \cite{DS,LPW,BHLN}. The difference
in constant terms is caused by that we used HQET for $T^{\mu}$, while full QCD
was used to calculate it in \cite{DS,LPW,BHLN}.
\par
The TMD light-cone wave function has a factorized relation to $\Phi_+$
in b-space\cite{MW}. It reads:
\begin{equation}
\phi_+(k^+, b, \zeta, \mu) = \int_0^{\infty} d q^+ C_B (k^+, q^+, b, \zeta, \mu) \Phi_+(q^+, \mu)
   +{\mathcal O}( b),
\end{equation}
where the function $C_B$ can be determined by perturbative theory and is free from any soft divergence.
At leading order of $\alpha_s$ the function $C_B(k^+, q^+, b, \zeta, \mu)$ is $\delta(k^+- q^+)$.
The result of $C_B$ at one-loop level can be found in \cite{MW}.
It should be noted that from the results in Sect.2.
the TMD light-cone wave function $\phi_+(k^+, k_\perp, \zeta)$  at one loop order in the momentum space contains
various infrared divergences. Some of them are proportional to the tree-level result, i.e.,
to $\delta^2 (\vec q_\perp)$,
some of them take a form like $1/(q^2_\perp +\lambda^2)$. These singularities do not cancel
if $\vec q_\perp$ goes to zero. But, when we transform $\phi_+(k^+, k_\perp, \zeta)$
into the $b$-space, i.e., when we integrate over $k_\perp$, some of these singularities
are cancelled, the remaining singularities are just the same as those in $\Phi_+$. Therefore
$C_B$ is free from any soft divergences. The same also happens to the soft factor $\tilde S$
with the difference that the infrared singularities are completely cancelled,
if we transform it into the $b$-space, or we integrate over the transverse momentum.
The soft factor $\tilde S$
in $b$-space reads:
\begin{eqnarray}
\tilde S (q^+, \vec b, \zeta_u, \mu) &=& \delta(q^+)
+ \frac{4\alpha_s}{3\pi} \theta (q^+)\left ( \frac{\ln (\tilde b^2 \zeta_v^2(q^+)^2)}{q^+}\right )_+
\nonumber\\
  && + \frac{2\alpha_s}{3\pi} \delta (q^+) \left [
           \ln(\tilde b^2 \mu^2) \left ( \ln\frac{\zeta_u^2}{\zeta_v^2} -1\right )
           +\frac{\pi^2}{6} +\frac{1}{2}\ln^2 (\tilde b^2 (P^+)^2 \zeta_v^2) \right ]
            +{\mathcal O}(b^2),
\end{eqnarray}
with $\tilde b = be^{\gamma}/2$. Here $\tilde S (q^+, \vec b, \zeta_u, \mu)$ should be taken
as a distribution for $q^+ < P^+$. The heavy quark limit implies $P^+ \to \infty$.
As discussed before and in \cite{MW}, we should take finite $P^+$ in the calculation
and take the limit in the final result. The same also applies for Eq.(43), where the upper bound of $q^+$
should be taken as $P^+$. With these results our factorization formula can be re-written as:
\begin{eqnarray}
F_V = F_A &=&\frac{i Q_q}{2 v\cdot p} \lim_{b\to 0}\lim_{ P^+\to \infty} \int_0^{P^+} d q^+  \Phi_+(q^+, \mu) \left \{
 H^{(0)} (q^+, \zeta_u, \mu) + H^{(1)}(q^+, \zeta_u, \mu)
\right.
\nonumber\\
   && \left. +\int_0^{P^+} dk^+ C_B^{(1)} (k^+, q^+,\zeta,\mu)H^{(0)} (l^+, \zeta_u, \mu)
\right.
\nonumber\\
        && \left. +\int_{-q^+}^{P^+}  d l^+ \tilde S^{(1)} ( l^+, b, \zeta_u,\mu)H^{(0)} (l^+ +q^+, \zeta_u, \mu)
  \right \} + {\mathcal O}(\alpha_s^2).
\end{eqnarray}
With our results of $C_B^{(1)}$, $\tilde S^{(1)}$ and $H^{(1)}$ we reproduce $H_c$ in Eq.(42). Therefore,
the two factorizations with fixed orders of perturbative theory are equivalent.

\par\vskip20pt
\noindent
{\bf\large 6. Resummation of Large Logarithms}
\par\vskip15pt
In general, one expects that the most important $k^+$-region of $\phi^+ (k^+, k_\perp,\zeta)$
for a convolution with the wave function like Eq.(37) will be around $k^+ \sim \Lambda_{QCD}$.
Also the important region of the soft factor $\tilde S(l^+,l_\perp, \zeta_u)$ is with small
$l^+$, i.e., $l^+\sim \Lambda_{QCD}$. This results in that $H^{(1)}(k^+ +l^+)$ will contain
large single logarithms and large double logarithms and it spoils the perturbative expansion
of $H$. Those large logarithms should be resummed for a reliable prediction.
\par
In the collinear factorization in \cite{DS,LPW,BHLN}, the resummation can be done by introducing
a jet factor in the frame work of the soft collinear effective theory\cite{SCET}, or a jet factor in the
full QCD\cite{Li3}.  Similarly,
we can also introduce a jet factor in our factorization for the resummation. However, as we have seen
before, our TMD light-cone wave function and soft factor depend on the parameter $\zeta_u$. This dependence
can be used to resum those large logarithms. Before showing this, let us first study the evolution
of the soft factor.
\par
The evolution with the renormalization $\mu$ and with the parameter $\zeta_u$ reads:
\begin{eqnarray}
  \mu \frac{\partial}{\partial \mu} \tilde S (k^+, \vec k_\perp, \zeta_u,\mu) &=&
  \frac{4\alpha_s}{3\pi} \left [ \ln\frac{\zeta^2_u}{\zeta^2_v} -1 \right ]\tilde S (k^+, \vec k_\perp, \zeta_u,\mu)
   +{\mathcal O}(\alpha_s^2),
\nonumber\\
\zeta_u \frac{\partial }{\partial \zeta_u}\tilde S (k^+, b, \zeta_u,\mu)  &=& \frac{4 \alpha_s}{3\pi}
\left [ 2 \gamma -\mbox{ln}4 +\mbox{ln} b^2 \mu^2 \right ] \tilde S (k^+, b, \zeta_u,\mu) +{\mathcal O}(\alpha_s^2),
\nonumber\\
\mu \frac{\partial }{\partial \mu} \phi_+ (k^+, k_\perp, \zeta, \mu )
 &=& \frac{\alpha_s}{\pi} \left [ 1 + \frac{2}{3} \left ( 2 -\ln\frac{\zeta_u^2}{\zeta_v^2} \right ) \right ]
        \phi_+ (k^+, k_\perp, \zeta, \mu ) +{\mathcal O}(\alpha_s^2),
\nonumber\\
\zeta \frac{\partial}{\partial \zeta} \phi_+(k^+,b, \zeta,\mu) &=&
   \left [-\frac{4\alpha_s}{3\pi} \ln\frac{\zeta^2 b^2 e^{2\gamma-1}}{4}
       -\frac{2\alpha_s}{3\pi} \ln\frac{\mu^2e}{\zeta^2} \right ]
       \phi_+(k^+,b,\zeta,\mu) +{\mathcal O}(\alpha_s^2),
\end{eqnarray}
where we also include the evolutions of the wave function for completeness.
With these equations, one can show that the form factors in Eq.(21) or Eq.(37)
are independent of $\zeta^2_u$, as expected. Also, their $\mu$-dependence
is compensated by the $\mu$-dependence of $f(\mu)$ in Eq.(15) so that
the matrix element
$\langle \gamma(\epsilon^*, p) \vert \bar q  \gamma^\mu (1-\gamma_5) b \vert \bar B(v)\rangle$
does not depend on $\mu$.
\par
To resum the large log terms,  we first take an initial value $\zeta_u =\zeta_{u 0}$ in Eq.(37) so that
there are no large log terms introduced by $\zeta_{u 0}$
in the wave function and the soft factor.
Then there will be large log terms with $\zeta_{u 0}$ in the coefficient function $H$, which
can be re-expressed with little algebra as:
\begin{eqnarray}
H(q^+, \zeta_{u0}, \mu) &=& \frac{1}{q^+} \left \{ 1 +
  \frac{2\alpha_s}{3\pi} \left[ \frac{1}{4} \ln^2 \left (\frac{\zeta^2_{u0}(q^+)^2}{e \mu^2}\right )
    -\frac{3}{2} \ln^2 \left(\frac{\mu}{q^+}\left ( \frac{\zeta_v^4\mu}{2p^-}\right )^{-\frac{1}{3}}  \right )
\right.\right.
\nonumber\\
    &&\left.\left. -\frac{1}{3} \left ( \ln\frac{2p\cdot v}{\mu} -\frac{3}{2} \right )^2
     - \frac{5\pi^2}{6} \right ] \right \} ,
\end{eqnarray}
where $q^+ = k^+ +l^+$.
For small $k^+$ and $l^+$ there are large log terms in the first line. These terms can be resummed by using
the $\zeta_u$-evolution of $H$:
\begin{equation}
\zeta_u \frac{\partial}{\partial \zeta_u} H(q^+, \zeta_u, \mu) =
\frac{2\alpha_s}{3\pi}\left [ \ln\frac{\zeta_u^2 (q^+)^2}{\mu^2} -1 \right ] H(q^+, \zeta_u, \mu) + {\mathcal O}(\alpha_s^2).
\end{equation}
Solving this equation we have:
\begin{eqnarray}
H(q^+, \zeta_{u0}, \mu) &=& \exp\left\{ +\frac{\alpha_s (\mu)}{6\pi}
        \left [
        \ln^2 \left (\frac{\mu^2 e }{\zeta^2_{u0} (q^+)^2}\right )
        - \frac{2}{3}\ln^2 \left (\frac{\zeta^4_v (q^+)^4}{2 p^-(q^+)\mu^2 }   \right ) \right ]
  \right \}
\nonumber\\
  && \cdot \frac{1}{q^+} \left \{ 1 +\frac{2\alpha_s(\mu) }{3\pi} \left[
       -\frac{1}{3} \left ( \ln\frac{2p\cdot v}{\mu} -\frac{3}{2} \right )^2
     - \frac{5\pi^2}{6} \right ] \right \}.
\end{eqnarray}
All large log terms due to small $+$-momenta are now resummed in the exponential.
To eliminate the large log terms in the second line of the above equation,
we can set $\mu = \mu_H$ with a large $\mu_H$ so that the $\ln(2p\cdot v/\mu_H)$ is a number of order 1.
Then there are large log terms due to large $\mu_H$
in the wave function and the soft factor. With the evolution equations in Eq.(46), we can evolute
them to lower scales as $\mu =\mu_0$.
Finally we have for the form factors:
\begin{eqnarray}
F_V = F_A &=&\frac{i Q_q}{2 v\cdot p} \lim_{b\to 0} \int dk^+  d l^+  \phi_+(k^+,b, k^+ \zeta_{u0}, \mu_0)
         \tilde S (l^+, b, \zeta_{u0}, \mu_0) \theta (k^+ +l^+)
\nonumber\\
         && \cdot  e^{S_F(k^+ +l^+)} \cdot \frac{1}{k^+ +l^+}
           \left \{ 1 +
\frac{2\alpha_s (\mu_H)}{3\pi} \left [-\frac{1}{3} \left ( \ln\frac{2p\cdot v}{\mu_H} -\frac{3}{2} \right )^2
     - \frac{5\pi^2}{6} \right ] \right\},
\end{eqnarray}
with
\begin{eqnarray}
S_F(q^+) &=& \frac{\alpha_s (\mu_H)}{6\pi}
        \left [
        \ln^2 \left (\frac{\mu_H^2 e }{\zeta^2_{u0} (q^+)^2}\right )
        - \frac{2}{3}\ln^2 \left (\frac{\zeta^4_v (q^+)^4}{2 p^-(q^+)\mu_H^2 }   \right )\right ]
\nonumber\\
     &&     +\int^{\mu_H}_{\mu_0} \frac{d\mu}{\mu} \frac{\alpha_s(\mu)}{\pi}
     \left ( 1 +\frac{2}{3} \ln\frac{\zeta^2_{u0}}{\zeta^2_v} \right ).
\end{eqnarray}
In the above all large logs are resummed in the factor $e^{S_F}$.
The initial value $\mu_0$ should be taken where perturbative QCD is still applicable.
One may take $\mu_0 = 1\sim 2$GeV. For $\zeta_{u0}$, with our definitions of the TMD light-cone
wave function and the soft factor, we should have $\zeta_{u0} >>1$, although the physics here, i.e.,
the form factors, does not depend on $\zeta_{u0}$. However, one should not take a too large $\zeta_{u0}$
to avoid large log terms in the wave function and the soft factor. A detailed study of a reasonable choice of
$\zeta_{u0}$ and $\mu_0$ is needed when one uses the factorization results for phenomenological applications.
\par
For the resummed results one can also use the relation in Eq.(43) and the result in Eq.(44) to
express them in term of the usual light-cone wave function $\Phi_+$, as in the last section.
Then instead of the integrand $C_B^{(1)}\otimes H^{(0)}$ in Eq.(45) we have a complicated integrand
$C_B^{(1)}\otimes H^{(0)}\otimes e^{S_F}$.
Unfortunately, we are unable to calculate the integral
analytically. The same also applies to the term corresponding to the term in the third line of Eq.(45).
Here, we only remind that our resummed form factors can be expressed as a convolution
of $\Phi_+$ with other functions.
\par\vskip20pt
\noindent
{\bf\large 7. Conclusion and Outlook}
\par\vskip15pt
As mentioned in the introduction, there are two approaches
for exclusive B-decays.
The two approaches are not only different in their formulations
but also in some predictions in comparison with experiment. This
leads to controversial discussions, e.g., see \cite{LiLiao,DS,
LN1,DS1}. Since two approaches are from one fundamental
theory--QCD, there must be some relations between them. With a
consistent definition of TMD light-cone wave functions these relations
can be explored and predictions for exclusive $B$-decays from the two approaches may be
unified. For this purpose, a first step is
to define the TMD light-cone wave function consistently and to
obtain relations between the TMD light-cone wave function  and the usual light-cone wave
function in the collinear factorization.
This has been done in our previous work\cite{MW}.
\par
In this paper,  we have shown  that with the consistent definition of
the TMD light-cone wave function the TMD factorization for
the radiative leptonic B-decay can be performed consistently at one-loop level.
In this factorization, beside the wave function as a nonperturbative object,
another nonperturbative object, which is the soft factor, must be introduced,
so that the perturbative coefficient function is free from any soft divergence.
The results are given in Eq.(37) and Eq.(38).
An extension of our factorization beyond one-loop level is possible.
\par
The TMD light-cone wave function defined in \cite{MW} does not only depend on $+$- and
transverse components of parton momentum, but also depends on the parameter
$\zeta_u$ which regularizes the light-cone singularity. This $\zeta_u$-dependence
can be used to resum large logarithms in the perturbative coefficient function,
as we have shown in Sect.6.
\par
For the decay studied here, we can show that the result of collinear factorization
can be derived from that of our TMD factorization. Hence the two factorizations are
related to each other. This simple relation is to be expected because there is no
hadrons, or partons in the final state. In other cases, the relation can be expected
to be complicated.
\par
Having shown that TMD factorization works in the simple case, we are ready to explore
how TMD factorization works in other complicated cases and how it is related to the
collinear factorization in these cases. Works for this are in progress.
\par\vskip20pt
\par\noindent
{\bf\large Acknowledgments}
\par
This work is supported by National Nature
Science Foundation of P.R. China.
\par\vskip40pt


\end{document}